\documentclass [12pt]{article}
\usepackage{amsmath}
\usepackage{epsfig}
\begin{document}
\def\b{\bar}
\def\d{\partial}
\def\D{\Delta}
\def\cD{{\cal D}}
\def\cK{{\cal K}}
\def\f{\varphi}
\def\g{\gamma}
\def\G{\Gamma}
\def\l{\lambda}
\def\L{\Lambda}
\def\M{{\Cal M}}
\def\m{\mu}
\def\n{\nu}
\def\p{\psi}
\def\q{\b q}
\def\r{\rho}
\def\t{\tau}
\def\x{\phi}
\def\X{\~\xi}
\def\~{\widetilde}
\def\h{\eta}
\def\bZ{\bar Z}
\def\cY{\bar Y}
\def\bY3{\bar Y_{,3}}
\def\Y3{Y_{,3}}
\def\z{\zeta}
\def\Z{{\b\zeta}}
\def\Y{{\bar Y}}
\def\cZ{{\bar Z}}
\def\`{\dot}
\def\be{\begin{equation}}
\def\ee{\end{equation}}
\def\bea{\begin{eqnarray}}
\def\eea{\end{eqnarray}}
\def\half{\frac{1}{2}}
\def\fn{\footnote}
\def\bh{black hole \ }
\def\cL{{\cal L}}
\def\cH{{\cal H}}
\def\cF{{\cal F}}
\def\cP{{\cal P}}
\def\cM{{\cal M}}
\def\ik{ik}
\def\mn{{\mu\nu}}
\def\a{\alpha}

\title{Instability of  Black Hole Horizon With Respect\\ to Electromagnetic
Excitations.}

\author{Alexander Burinskii \\
 NSI, Russian Academy of Sciences}

\date{First Award in the 2009 Essay Competition of the Gravity Research Foundation.
} \maketitle

\begin{abstract}
Analyzing  exact solutions of the Einstein-Maxwell equations in the
Kerr-Schild formalism we show that black hole horizon is instable
with respect to electromagnetic excitations. Contrary to perturbative
smooth harmonic solutions, the exact solutions for electromagnetic
excitations on the Kerr background are accompanied by singular beams
which have very strong back reaction to metric and break the horizon,
forming the holes which allow radiation to escape interior of black-hole.
As a result, even the weak vacuum fluctuations break the horizon
topologically, covering it by a set of fluctuating microholes. We
conclude with a series of nontrivial consequences, one of which is
that there is no information loss inside of black-hole.
\end{abstract}

\newpage

 The statements on stability of the black-hole horizon
are based on the theorems (Robinson and Carter) claiming the
uniqueness of the Kerr solution. As usual, these theorems are
valid under a series of conditions, among which are stationarity,
axial symmetry and asymptotic flatness of the metric
\cite{Chandra}. Meanwhile, many of these conditions turns out to
be vulnerable in practice. In
particular, in the problem of interaction of a
black-hole (BH) with electromagnetic perturbations, the
conditions of stationarity and axial symmetry are broken, as well
as the condition of asymptotic flatness, since the perturbations are
coming from asymptotic region and are scattered back to infinity.

Second vulnerable point is the use of perturbative approach.
The BH solutions belong to type D which has two principal
null congruences (PNC) $\cal K^{\pm} .$ All the tensor fields of
the exact solutions of type D are to be  aligned to the PNC.
Meanwhile, the Kerr and Kerr-Newman (KN) solutions are twosheeted,
and $\cal K^{+} $ (PNC on the $(+) $ sheet) differs from  $\cal K^{-}$
(PNC on the $(-) $ sheet), which is exhibited in the Kerr-Schild
 form of metric
\be g_\mn^\pm =\eta_\mn + 2H k_\m^\pm k_\n^\pm, \label{KS}\ee were
$\eta_\mn$ is metric of auxiliary Minkowski space-time $M^4$ and
$k^{\m(x)\pm}, \ (x=x^\m \in M^4)$ are two different vector fields
tangent to the corresponding $\cal K^{\pm} .$ Direction
of PNC is determined in the null coordinates $u=z-t,\quad
v=z+t,\quad\zeta=x+iy,\quad\bar\zeta=x-iy,$ by 1-form \be
k_\m^{(\pm)} dx^\m = P^{-1}(du +\Y^\pm d\z + Y^\pm d\Z - Y^\pm
\Y^{(\pm)} dv), \label{kpm}\ee and depends on the  complex
coordinate on celestial sphere \be Y=e^{i\phi} \tan \frac
{\theta}{2}  \label{Y} .\ee

Two solutions $Y^\pm(x)$ are determined by the Kerr Theorem
\cite{DKS,Multiks,BurKer,BurNst,KraSte}.

 Projection of the Kerr PNC on auxiliary background $M^4 $ is
 depicted on Fig.1  .

\begin{figure}[ht]
\centerline{\epsfig{figure=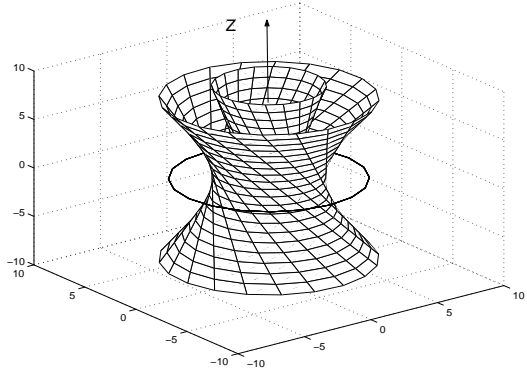,height=5cm,width=7cm}}
\caption{The Kerr singular ring and congruence. }
\end{figure}

The PNC covers the spacetime twice: in the form of ingoing PNC
$k^{\m -} \in \cal K^- $ which falls on the disk spanned by Kerr
singular ring, and as outgoing one, $k^{\m +} \in \cal K^+ ,$
positioned on the second sheet of the same spacetime $M^4 .$ The
metric $g_\mn^\pm =\eta_\mn + 2H k_\m^\pm k_\n^\pm$ and
electromagnetic fields, being aligned with PNC, are to be different on
the in- and out- sheets and should not be mixed. This
twosheetedness is ignored in perturbative approaches, leading to
drastic discrepancies in the solutions and  in the structure of
horizon. {\it The typical exact electromagnetic solutions on the
Kerr background take the form of singular beams propagating along
the rays of PNC, contrary to smooth angular dependence of the wave
solutions used in perturbative approach!}

Position of the horizon is determined
by function $H $ which has for the exact KS solutions the form,
\cite{DKS}, \be H =\frac {mr - |\psi|^2/2} {r^2+ a^2 \cos^2\theta}
 \ , \label{Hpsi} \ee where $\psi(x)$ is related to vector potential
of electromagnetic field

\be \alpha =\alpha _\m dx^\m \\
= -\frac 12 Re \ [(\frac \psi {r+ia \cos \theta}) e^3 + \chi d \Y
], \quad \chi = 2\int (1+Y\Y)^{-2} \psi dY  \ , \label{alpha} \ee
 which obeys
the alignment condition \be \alpha _\m k^\m=0 . \ee The equations
(\ref{KS})and (\ref{Hpsi}) display compliance and elasticity of
the horizon with respect to electromagnetic field.

The Kerr-Newman solution corresponds to $\psi=q=const.$. However,
any nonconstant holomorphic function $\psi(Y) $ yields also an
exact KS solution, \cite{DKS}. On the other hand, any nonconstant
holomorphic functions on sphere acquire at least one pole.
A single pole at $Y=Y_i$
\be \psi_i(Y) = q_i/(Y-Y_i) \ee produces the beam in angular
directions \be Y_i=e^{i\phi_i} \tan \frac {\theta_i}{2} \label{Yi}
.\ee

The function $\psi(Y)$ acts immediately on the function
$H$ which determines the metric and the position of the horizon.
 The given in \cite{BEHM1} analysis showed
 that electromagnetic beams have very strong back reaction to metric
 and deform topologically the horizon, forming the holes which
allows matter to escape interior (see fig.2).

\begin{figure}[ht]
\centerline{\epsfig{figure=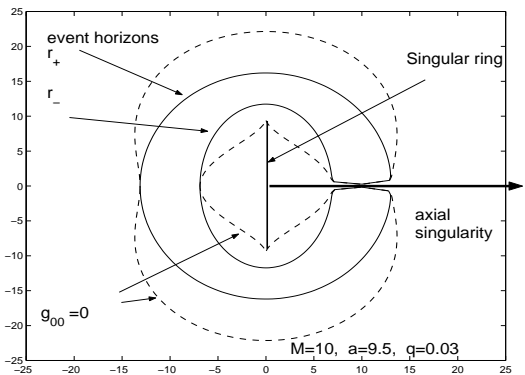,height=5cm,width=7cm}}
\caption{Hole in the horizon of a rotating black hole formed by a
singular beam directed along axis of symmetry.}
\end{figure}

The exact KS solutions may have arbitrary number of beams in
angular directions $Y_i=e^{i\phi_i} \tan \frac {\theta_i}{2}.$
The corresponding function
\be \psi (Y) = \sum _i \frac {q_i} {Y-Y_i},
\label{psiY}\ee leads to the horizon with many holes.
 In the far zone the beams tend to the
known exact singular pp-wave solutions (A.Peres) \cite{KraSte}. The
multi-center KS solutions considered in \cite{Multiks} showed that
the beams are extended up to other matter sources, which may
also be assumed at infinity.

The stationary KS beamlike solutions may be generalized to
time-dependent wave pulses, \cite{BurAxi}, which tend to exact
 solutions in the low-frequency limit \cite{EM3p}.

Since the horizon is extra sensitive to electromagnetic
excitations, it may also be sensitive to the vacuum
electromagnetic field which is exhibited classically as a Casimir
effect, and it was proposed in \cite{EM3p,BEHM2} that the vacuum
beam pulses shall produce a fine-grained structure of fluctuating
microholes in the horizon, allowing radiation to escape interior
of black-hole, as it is depicted on Fig.3.

\begin{figure}[ht]
\centerline{\epsfig{figure=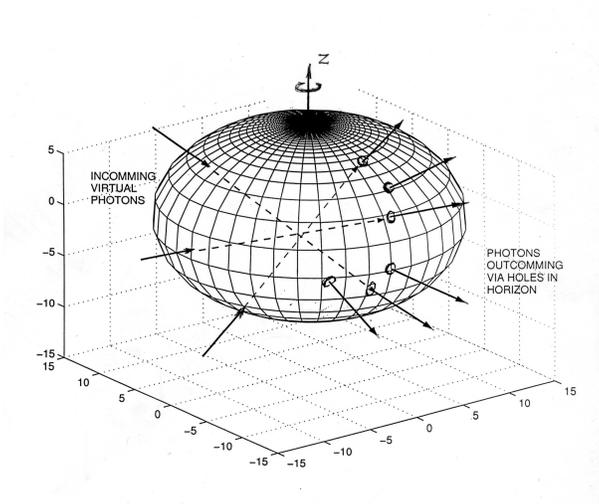,height=7cm,width=8cm}}
\vspace*{-7mm} \caption{Excitation of a black hole by the
zero-point field of virtual photons forming a set of micro-holes
at its horizon.}
\end{figure}

The function $\psi(Y,\t)$ corresponding to beam pulses has to
depend on retarded time $\t $ and satisfy to the obtained in
\cite{DKS} nonstationary Debney-Kerr-Schild (DKS) equations
leading to the extra long-range radiative term $\gamma(Y,\t)Z .$
The expression for the null electromagnetic radiation take the
form, \cite{DKS}, $F^\mn = Re \cF _{31} e^{3\m}\wedge e^{1\n},$
where \be  \cF _{31}=\gamma Z - (AZ),_1 \ , \ee  $Z=P/(r+ia\cos
\theta), \quad P=2^{-1/2}(1+\Y\bar \Y) ,$ and the null tetrad
vectors have the form $e^{3\m}= Pk^\m, \ e^{1\m} = \d_\Y e^{3\m}.$

The long-term attack on the DKS equations performed in
\cite{BurKer,BurNst,BurAxi,EM3p,BEHM2} has led to the obtained in
\cite{HolEvap} time-dependent solutions which revealed  a
holographic structure of the fluctuating Kerr-Schild spacetimes
and showed explicitly that electromagnetic radiation from a
black-hole interacting with vacuum contains two components:

a) a set of the singular beam pulses (determined by function
$\psi(Y,\t) ,$) propagating along the Kerr PNC and breaking the
topology and stability of the horizon;

b)the regularized radiative component (determined by
$\gamma_{reg}(Y,\t)$) which is smooth and,  similar to that of the
the Vaidya `shining star' solution \cite{KraSte}, determines
evaporation of the black-hole, \be\dot m = - \frac 12
P^2<\gamma_{reg}\bar \gamma_{reg}> .\ee

The mysterious twosheetedness of the Kerr-Schild geometry plays
principal role in the holographic black-hole spacetime
\cite{HolEvap}, allowing one to consider action of the
electromagnetic in-going vacuum as a time-dependent process of
scattering. The obtained solutions describe excitations of
electromagnetic beams on the Kerr-Schild background, the
fine-grained fluctuations of the black-hole horizon, and the
consistent back reaction of the beams to metric. The holographic
space-time is twosheeted and forms a fluctuating pre-geometry
which reflects the dynamics of singular beam pulses. This
pre-geometry is classical, but  has to be still regularized to get
the usual smooth classical space-time. In this sense it takes an
intermediate position between the classical and quantum gravity.

We arrive at the following nontrivial conclusions:

\begin{itemize}

\item contrary to the perturbative results, the exact
time-dependent solutions of the Maxwell equations on the Kerr
background  contain commonly the light-like singular beam pulses
which lead unavoidable to formation of the holes in horizon,

\item topological instability of black hole horizon with respect to
electromagnetic excitations rejects the recent speculations on the
possible creation of black holes in the scattering processes at
high energies.

\item black hole horizon is topologically nonstable even
with the respect to the very weak electromagnetic excitations, in
particular, with respect to the action of vacuum fluctuations
leading to the fine-grained topological fluctuations of the
horizon structure,

\item interior of a black hole is not isolated from the
exterior region and matter may leave the interior in
the form of the outgoing null radiation, and so, there is no
information loss paradox.

\end{itemize}

\end{document}